\documentclass[11pt]{article}

\usepackage{amsmath}
\usepackage{graphicx}
\usepackage{indentfirst}
\usepackage{amssymb}
\usepackage{cite}
\usepackage{color}
\usepackage{subfigure}

\begin{document}

\title{\bf{Weak Cosmic Censorship in Kerr-Sen Black Hole\\under Charged Scalar Field}}

\date{}
\maketitle

\begin{center}
\author{Bogeun Gwak}$^a$\footnote{rasenis@dongguk.edu}\\

\vskip 0.25in
$^{a}$\it{Division of Physics and Semiconductor Science, Dongguk University, Seoul 04620,\\Republic of Korea}\\
\end{center}
\vskip 0.6in

{\abstract
{We investigate the weak cosmic censorship conjecture for Kerr--Sen black holes, which are solutions to the four-dimensional low-energy effective field theory for the heterotic string theory, based on the scattering of a charged scalar field. When the fluxes of the scalar field are assumed to transfer its conserved quantities to the black hole, extremal and near-extremal black holes cannot be over-spun and over-charged in their first-order variations, which is sufficient to conclude that the weak cosmic censorship conjecture is valid for Kerr--Sen black holes. We confirm our conclusion by relating it to the first, second, and third laws of thermodynamics.}}

\thispagestyle{empty}
\newpage
\setcounter{page}{1}

\section{Introduction}\label{sec1}

The black hole, one of the compact cosmic objects, has an event horizon. Classically, matter passing through the horizon cannot escape to the outside of the horizon. Therefore, an observer cannot detect any energy from the inside of a black hole, which is one of reasons why black holes are interesting objects. The energy of a black hole is divided into irreducible mass and reducible energy\cite{Christodoulou:1970wf,Bardeen:1970zz,Christodoulou:1972kt}. The irreducible mass increases with any classical process, but the reducible energy is different. Rotational and electric energies are examples of reducible energies that are reduced or extracted by specific processes such as the Penrose process\cite{Penrose:1971uk}. However, when a quantum process is considered at the horizon, black holes are found to emit radiation; therefore, black holes are considered to be thermodynamic objects with a Hawking temperature\cite{Hawking:1974sw,Hawking:1976de}. Furthermore, according to the increase in the irreducible mass of a black hole, the Bekenstein--Hawking entropy can be defined to be proportional to the surface area of the black hole\cite{Bekenstein:1973ur,Bekenstein:1974ax}. The laws of thermodynamics are accordingly constructed by these thermodynamic variables.

The event horizon plays a role in hiding the inner structure of black holes, which is quite important to the curvature singularity. A naked singularity, which is exposed to an observer without the horizon, causes a breakdown of causality. However, this cannot happen, because  the singularity is hidden by the horizon. This conjecture is called the weak cosmic censorship conjecture (WCCC)\cite{Penrose:1964wq,Penrose:1969pc}. In the WCCC, the horizon should be hidden inside the black hole to prevent a naked singularity in the universe. Since the test for Kerr black holes based on adding a particle was proposed\cite{Wald:1974ge}, various investigations of the WCCC of numerous black holes have been reported. Because the validity of WCCC depends on the testing channels, the conclusions can be inconsistent. Particularly, when tested by adding a particle, the WCCC of Reissner--Nordstr\"{o}m black holes was found to be invalid\cite{Hubeny:1998ga}, but another study found that the WCCC of the black hole remains valid\cite{Isoyama:2011ea}. Furthermore, the states of black holes affect the validity of the WCCC. The WCCC of the extremal Kerr black hole is known to be valid. However, the WCCC of the near-extremal case has been found to be invalid \cite{Jacobson:2009kt}, but in a different analysis, the invalidity was resolved by considering the self-force or back-reaction\cite{Barausse:2010ka,Barausse:2011vx,Colleoni:2015ena,Colleoni:2015afa,Sorce:2017dst}. The WCCC is still actively studied in many black holes by using various methods\cite{Gao:2012ca,Zhang:2013tba,Rocha:2014jma,McInnes:2015vga,Gwak:2015fsa,Cardoso:2015xtj,Siahaan:2015ljs,Gwak:2016gwj,Horowitz:2016ezu,Revelar:2017sem,Song:2017mdx,Yu:2018eqq,Liang:2018wzd,Gwak:2018tmy,Zeng:2019jrh,Han:2019kjr,Zeng:2019jta,Chen:2019pdj,Han:2019lfs,Zeng:2019aao,Chen:2019nsr,Wang:2019dzl,Zeng:2019hux,Shaymatov:2019upj,Hong:2019yiz,Mu:2019bim,Zeng:2019baw,Shaymatov:2019pmn,Wang:2019jzz,He:2019kws,Hu:2019zxr}.  In particular, the WCCC can be investigated based on the scattering of a scalar field, where changes in the black hole depend on the modes of the scalar field\cite{Hod:2008zza,Semiz:2005gs,Toth:2011ab,Natario:2016bay,Duztas:2017lxk,Duztas:2018adf,Gwak:2019asi,Duztas:2019ick,Natario:2019iex,Jiang:2019vww,Gwak:2020zht}. Moreover, the boundary condition of the scalar field differs according to the sign of the cosmological constant in Einstein's gravity. Nevertheless, the WCCC was found to be valid in a unified description of the arbitrary cosmological constant\cite{Gwak:2018akg}.

In this work, we investigate the WCCC in Kerr--Sen black holes based on the scattering of a {\it{charged}} scalar field. The Kerr--Sen black hole is a solution to the four-dimensional low-energy effective field theory for the heterotic string theory\cite{Sen:1992ua}. Because the Kerr--Sen black hole involves a combination of Maxwell and antisymmetric tensor fields, its electric potential is different from that of the Reissner--Nordstr\"{o}m black hole. Thus, the Kerr--Sen black hole is an interesting case to investigate whether the WCCC remains valid in a black hole based on the low-energy effective field theory for the heterotic string theory. Herein, we introduce the charged scalar field, which plays an important role in the validity of the WCCC. Because the electric potential is different from that of Reissner--Nordstr\"{o}m black holes, the scalar field should be coupled with the rotational and electric contents of the black hole. Note that, by particle absorption, the WCCC was previously found to be violated in the Kerr--Sen black hole \cite{Siahaan:2015ljs}, but the WCCC was found to be valid in \cite{Gwak:2016gwj} when considering the second-order variation owing to the charged particle. On the other hand, with a {\it{neutral}} scalar field, the violation of the WCCC has been shown in the first order\cite{Duztas:2018adf}. With the Iyer--Wald formalism\cite{Iyer:1994ys}, which is a different method, the WCCC remains invalid in the first-order perturbation, but the WCCC has been shown to be valid in \cite{Jiang:2019vww} by considering up to the second-order perturbation with charged matter. In the present study, with the actual scattering process of a charged scalar field, we investigate the evolution of the black hole according to the fluxes of the scalar field. Consequently, we prove that the WCCC is {\it{valid}} in the {\it{first-order variation}} of the black hole. Furthermore, we {\it{ensure}} that our conclusion regarding the validity of the WCCC is consistent with the first, second, and third laws of thermodynamics.

The remainder of this paper is organized as follows. In Sec.\,\ref{sec2}, we review Kerr--Sen black holes. In Sec.\,\ref{sec3}, the charged-scalar-field equations are solved at the outer horizon, and the fluxes of the scalar field are obtained. In Sec.\,\ref{sec4}, the WCCC is investigated in extremal and near-extremal black holes. In Sec.\,\ref{sec5}, we confirm our conclusion regarding the validity of the WCCC in terms of the laws of thermodynamics. Sec.\,\ref{sec6} concludes this study.

\section{Kerr--Sen Black Holes}\label{sec2}

The Kerr--Sen black hole is a solution to the four-dimensional low-energy effective field theory for the heterotic string theory. In the low-energy limit, the bosonic part of the heterotic string theory is given as the four-dimensional effective action in the string frame\cite{Sen:1992ua}.
\begin{align}\label{eq:lowenergyeffaction1}
S=-\int d^4 x \sqrt{-g}e^\Phi\left(-R+\frac{1}{12}H_{\mu\nu\rho}H^{\mu\nu\rho}-g^{\mu\nu}\partial_\mu\Phi\partial_\nu\Phi+\frac{1}{8}F_{\mu\nu}F^{\mu\nu}\right),
\end{align}
where $\Phi$ is the dilaton field. The Maxwell and 3-form field strengths, $F_{\mu\nu}$ and $H_{\mu\nu\rho}$, respectively, are combinations of the Maxwell and antisymmetric tensor fields denoted as $A_\mu$ and $B_{\mu\nu}$.
\begin{align}
H_{\mu\nu\rho}&=\partial_\mu B_{\nu\rho}+\partial_\rho B_{\mu\nu}+\partial_\nu B_{\rho\mu}-\frac{1}{4}\left(A_\mu F_{\nu\rho}+A_\rho F_{\mu\nu}+A_\nu F_{\rho\mu}\right),\quad F_{\mu\nu}=\partial_\mu A_{\nu}-\partial_\nu A_{\mu}.
\end{align}
By solving the equations of motion obtained from Eq.\,(\ref{eq:lowenergyeffaction1}), the charged black hole can be generated in the heterotic string theory by the Hassan--Sen transformation\cite{Hassan:1991mq}. The Kerr--Sen black hole, generated from the Kerr black hole to carry an electric charge, is in the Einstein frame
\begin{align}\label{eq:metric1}
ds^2&=-\frac{\Delta_r}{\rho_b^2}\left(dt-a\sin^2\theta d\phi\right)^2+\frac{\rho_b^2}{\Delta_r}dr^2+\rho_b^2 d\theta^2+\frac{\sin^2\theta}{\rho_b^2}\left(a\,dt-\left(r^2+a^2+2br\right)d\phi\right)^2,\\
\rho^2&=r^2+a^2\cos^2\theta,\quad\rho_b^2=\rho^2+2br,\quad\Delta_r=r^2+a^2-2(M-b)r,\quad b=\frac{Q^2}{2M},\quad J=Ma,\nonumber
\end{align}
which carries mass $M$, angular momentum $J$, and electric charge $Q$. Because the Kerr--Sen black hole is generated from the Kerr black hole, the rotational part of Eq.\,(\ref{eq:metric1}) coincides with that of the Kerr black hole, but the electric part is distinguishable according to the Hassan--Sen transformation. This relationship is also observed in components of the Maxwell field.
\begin{eqnarray}
 A_t=-\frac{Qr}{\rho_b^2}\,,\quad A_\phi=\frac{aQr\sin^2\theta}{\rho_b^2},
\end{eqnarray}
which are different from those of the Kerr--Newman black hole. Therefore, the effect of the electric charge is important to understand the Kerr--Sen black hole. There are two horizons in spacetime: the inner and outer horizons, which are, respectively, expressed as
\begin{align}
r_\text{i}=M-b-\sqrt{\left(M-b\right)^2-a^2},\quad r_\text{h}=M-b+\sqrt{\left(M-b\right)^2-a^2}.
\end{align}
The extremal condition and the location of the extremal horizon at this condition are, respectively,
\begin{align}
|a| \leq \left|M-b\right|,\quad r_\text{e}=M-b.
\end{align}
The angular velocity and surface area at the outer horizon explicitly depend on the charge and spin parameters.
\begin{align}
\Omega_\text{h}=\frac{a}{r_\text{h}^2+a^2+2br_\text{h}},\quad A_\text{h}=4\pi (r_\text{h}^2+a^2+2br_\text{h}).
\end{align} 
The Kerr--Sen black hole can be also considered as a thermodynamic system with a Hawking temperature. Hence, the laws of thermodynamics are applicable to the black hole. The first law of thermodynamics is expressed as follows.
\begin{align}
dM=T_\text{h} dS_\text{h}+\Omega_\text{h} dJ +\Phi_\text{h} dQ,
\end{align}
where the Hawking temperature, Bekenstein--Hawking entropy, and electric potential are, respectively, given as
\begin{align}
T_\text{h}=\frac{Q^2-2M(M-r_\text{h})}{8\pi M^2 r_\text{h}},\quad S_\text{h}=\frac{1}{4}A_\text{h}=\pi (r_\text{h}^2+a^2+2br_\text{h}),\quad \Phi_\text{h}=\frac{Q}{2M}.
\end{align}

\section{Fluxes of Charged Scalar Field}\label{sec3}

We herein consider the scattering of a massive scalar field with an electric charge by a Kerr--Sen black hole. With the scattering, the scalar field carries energy, angular momentum, and electric charge to the black hole. The conserved quantities of the black hole also vary as much as those of the scalar field. Hence, to estimate changes in the black hole, we need to find the amount of conserved quantities carried to the black hole through the outer horizon. These are given by the fluxes of the scalar field.

The action of a massive scalar field with an electric charge is expressed as
\begin{align}\label{eq:lagscalar01}
S_\Psi =-\frac{1}{2}\int d^D x \sqrt{-g}\left(\mathcal{D}_\mu \Psi \mathcal{D}^{*\mu} \Psi^*+\mu^2 \Psi \Psi^* \right),\quad \mathcal{D}_\mu=\partial_\mu-iq A_\mu,
\end{align}
where $q$ and $\mu$ are the electric charge and mass of the scalar field. Then, the equations of motion are
\begin{align}\label{eq:equationofmotion1}
\frac{1}{\sqrt{-g}}\mathcal{D}_\mu \left(\sqrt{-g} g^{\mu\nu} \mathcal{D} _\nu \Psi\right)-\mu^2 \Psi=0,\quad \mathcal{D}^*_\mu \left(\sqrt{-g} g^{\mu\nu} \mathcal{D}^* _{\nu} \Psi^*\right)-\mu^2 \Psi^*=0.
\end{align}
The equations of motion (\ref{eq:equationofmotion1}) are trivial in the separation of coordinates for $t$ and $\phi$. We can construct an ansatz such that
\begin{align}\label{eq:solution05}
\Psi(t,r,\theta,\phi)=e^{-i\omega t}R(r) \Theta(\theta) e^{im\phi}.
\end{align}
Then, the radial and $\theta$-directional equations are separable to
\begin{align}\label{eq:radialandthetadirectionaleq}
&\frac{1}{R}\partial_r \left(\Delta_r \partial _r R\right)+\frac{1}{\Delta_r}[-(r^2+a^2+2br)\omega+am+qQr]^2-\mu^2 (r^2 +2br)+2a\omega m-a^2\omega^2-\lambda=0,\\
&-\frac{1}{\Theta\sin\theta}\partial_\theta \left(\sin\theta \partial_\theta \Theta\right)+\mu^2 (a^2 \cos^2\theta)+(a\omega\sin\theta)^2 +(m\csc\theta)^2-a^2\omega^2-\lambda=0,\nonumber
\end{align}
where $\lambda$ is a separate variable. Here, the $\theta$-directional equation is the well-studied form of the scalar spheroidal harmonics\cite{Berti:2005gp,Gwak:2019ttv}. The scalar spheroidal harmonics is a generalization of the simple harmonics for a spin-zero field. Hence, we expect that there are solutions satisfying the $\theta$-directional equation in Eq.\,(\ref{eq:radialandthetadirectionaleq}). Then, the remaining part is the radial equation, which is key to analyze the fluxes of the scalar field. To solve the radial equation, we define a tortoise coordinate such as
\begin{align}
\frac{dr^*}{dr}=\frac{r^2+a^2+2br}{\Delta_r},
\end{align}
where the interval $(r_\text{h},\infty)$ in the radial coordinate transforms into $(-\infty,+\infty)$ in the tortoise coordinate. In the tortoise coordinate, the radial equation is given as  
\begin{align}\label{eq:radialeqfirst}
\frac{d^2 R}{d {r^*}^2}+\frac{2(r+b)\Delta_r}{(r^2+a^2+2br)^2}\frac{d R}{d {r^*}}&+\frac{1}{(r^2+a^2+2br)^2}[(r^2+a^2+2br)\omega-am-qQr]^2R\\&+\frac{\Delta_r}{(r^2+a^2+2br)^2}\left(-\mu^2 (r^2 +2br)+2a\omega m-a^2\omega^2-\lambda\right)R=0.\nonumber
\end{align}
By solving Eq.\,(\ref{eq:radialeqfirst}), we can obtain the radial waveform of the scalar field in all ranges. However, the energy, angular momentum, and electric charge of the scalar field are important here to estimate changes in the black hole. These can be described by the fluxes of the scalar field at the outer horizon. Hence, we need to solve for the fluxes of the scalar field at the outer horizon. In the tortoise coordinate, the radial equation in Eq.\,(\ref{eq:radialeqfirst}) reduces to a Schr\"{o}dinger-like equation at the outer horizon.
\begin{align}\label{eq:tortoiseradialeq02}
\frac{d^2 R}{d {r^*}^2}+\left(\omega-m\Omega_\text{h}-q\Phi_\text{h}\right)^2 R=0.
\end{align}
Therefore, the solutions to Eq.\,(\ref{eq:tortoiseradialeq02}) for an ingoing scalar field are obtained as
\begin{align}\label{eq:scalarsolution02}
\Psi=e^{-i\omega t} e^{- i \left(\omega - m\Omega_\text{h}-q\Phi_\text{h}\right)r^*} \Theta(\theta)e^{im\phi},\quad \Psi^*=e^{i\omega t} e^{ i \left(\omega - m\Omega_\text{h}-q\Phi_\text{h}\right)r^*} \Theta^*(\theta)e^{-im\phi}.
\end{align}
The fluxes of the scalar field in Eq.\,(\ref{eq:scalarsolution02}) represent the amount of energy, angular momentum, and electric charge flowing into the black hole. The fluxes are obtained from the energy-momentum tensor given by the Lagrangian in Eq.\,(\ref{eq:lagscalar01}). The energy-momentum tensor is expressed as
\begin{align}
T^\mu_\nu=\frac{1}{2}\mathcal{D}^\mu\Psi \partial_\nu \Psi^*+\frac{1}{2}\mathcal{D}^{*\mu}\Psi^* \partial_\nu \Psi-\delta^\mu_\nu\left(\frac{1}{2}\mathcal{D}_\mu\Psi \mathcal{D}^{*\mu}\Psi^* -\frac{1}{2}\mu^2\Psi\Psi^*\right).\nonumber
\end{align}
Then, the fluxes are
\begin{align}\label{eq:fluxes21}
&\frac{dE}{dt}=2\pi^2 M r_\text{h} \omega (\omega-m\Omega_\text{h}-q \Phi_\text{h}),\\
&\frac{de}{dt}=2\pi^2 M r_\text{h} q (\omega-m\Omega_\text{h}-q \Phi_\text{h}),\nonumber\\
&\frac{dJ}{dt}=2\pi^2 M r_\text{h} m (\omega-m\Omega_\text{h}-q \Phi_\text{h}).\nonumber
\end{align}
Because the fluxes in Eq.\,(\ref{eq:fluxes21}) show the rates of conserved quantities flowing into the black hole, we can expect that the change in the conserved quantities of the black hole is equal to the change in those of the ingoing scalar field. Then, during an infinitesimally small time interval, the conserved quantities of the black hole change as
\begin{align}\label{eq:variationbh01}
&dM=2\pi^2 M r_\text{h} \omega (\omega-m\Omega_\text{h}-q \Phi_\text{h})dt,\\
&dQ=2\pi^2 M r_\text{h} q (\omega-m\Omega_\text{h}-q \Phi_\text{h})dt,\nonumber\\
&dJ=2\pi^2 M r_\text{h} m (\omega-m\Omega_\text{h}-q \Phi_\text{h})dt.\nonumber
\end{align}
Here, superradiance, or the extraction of energy from a black hole, occurs when the frequency of the scalar field is less than the angular velocity or electric potential of the black hole. Explicitly,
\begin{align}\label{eq:super20aq}
\omega < m\Omega_\text{h}+q\Phi_\text{h}.
\end{align}
Thus, we consider the scattering of the scalar field including superradiance. Hence, the scalar field can increase or decrease the conserved quantities of the black hole under the scattering.

\section{Weak Cosmic Censorship in Kerr--Sen Black Holes}\label{sec4}

When the fluxes of the scalar field enter the Kerr--Sen black hole, the conserved quantities of the black hole vary according to Eq.\,(\ref{eq:variationbh01}). Because the variations of conserved quantities are infinitesimal during an infinitesimal time interval, the outer horizon of a non-extremal black hole remains stable under the scattering. However, the stability of the outer horizon of extremal and near-extremal black holes can be impacted because it is tightly balanced with the conserved quantities of the black holes under the extremal condition. Hence, with a small perturbation originating from the scalar field, extremal and near-extremal black holes have a possibility of being over-spun or over-charged. Consequently, the singularity will be exposed to an outside observer, making the WCCC invalid in this case. Here, we will investigate whether extremal and near-extremal black holes can be over-spun or over-charged by the conserved quantities carried by the fluxes of the scalar field during an infinitesimal time interval.

\subsection{Extremal Black Holes}\label{sec41}

An over-spun or over-charged black hole has no horizons. Therefore, its singularity is naked. This implies that $\Delta_r$ has no solution in Eq.\,(\ref{eq:metric1}), because solutions to $\Delta_r$ correspond to the inner and outer horizons of the Kerr--Sen black hole. Between the locations of two horizons, $\Delta_r$ has the minimum value at $r=r_\text{min}$. Particularly, the extremal black hole has one horizon located at the minimum value. In terms of $\Delta_r$, the extremal black hole satisfies 
\begin{align}\label{eq:extremalcond01}
r_\text{h}=r_\text{min},\quad \left.\Delta_\text{min}\equiv\Delta_r\right|_{r=r_\text{min}}=0,\quad \left.\frac{\partial \Delta_r}{\partial r}\right|_{r=r_\text{min}}\equiv \frac{\partial\Delta_\text{min}}{\partial r_\text{min}}=0,\quad \left.\frac{\partial^2 \Delta_r}{\partial r^2}\right|_{r=r_\text{min}}>0,
\end{align}
where the minimum value saturates to zero for an extremal black hole. When the scalar field carries conserved quantities to the extremal black hole, its mass, angular momentum, and electric charge are varied owing to the fluxes in Eq.\,(\ref{eq:variationbh01}). Furthermore, because $\Delta_r$ is a function of $(M,J,Q)$, the fluxes change $\Delta_r(M,J,Q)$ into $\Delta_r(M+dM,J+dJ,Q+dQ)$ during a time interval $dt$. Hence, the extremal black hole is slowly varied owing to the scalar field. Note that the fluxes of the scalar field are very small.
\begin{figure}[h]
\centering
\subfigure[{$\Delta_r$ of the extremal and near-extremal black holes.}] {\includegraphics[scale=0.90,keepaspectratio]{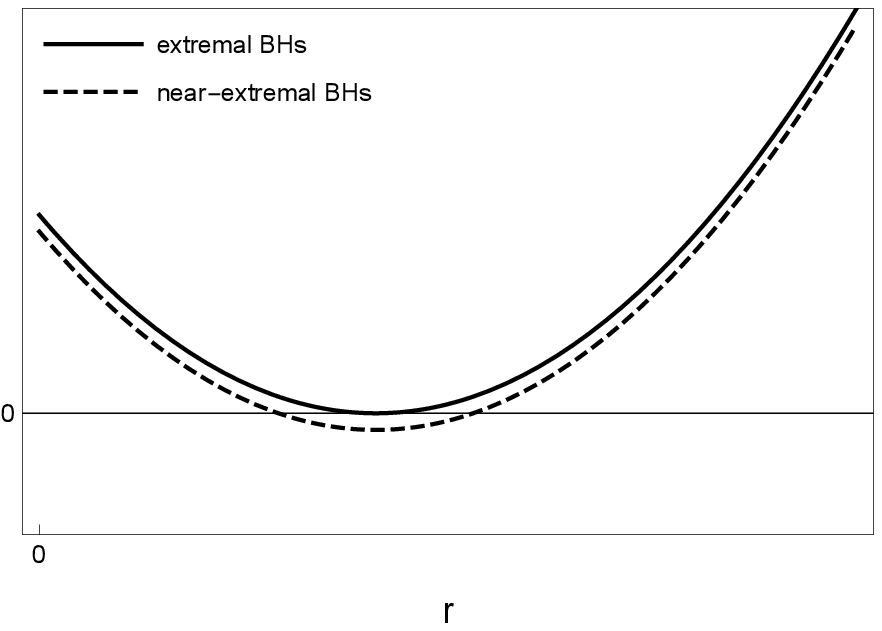}}\quad
\subfigure[{$\Delta_r$ of black holes in three possible final states.}] {\includegraphics[scale=0.90,keepaspectratio]{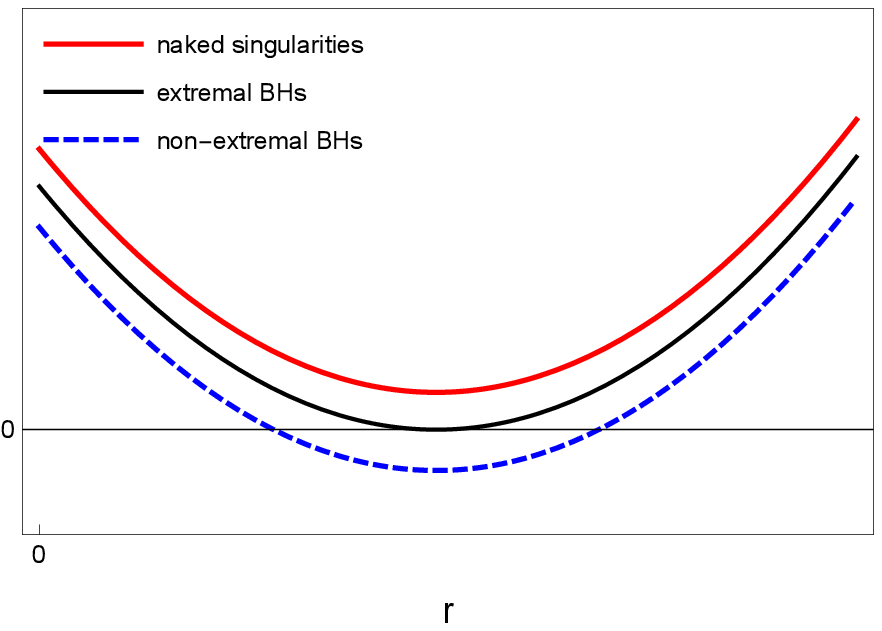}}
\caption{{\small Graphs of $\Delta_r$ in initial and final states.}}
\label{fig:fig1extremal}
\end{figure}
Therefore, we can assume that changes in the black hole are very slow during the time interval. The minimum value of $\Delta_\text{min}(M,J,Q)$ moves that of $\Delta_r(M+dM,J+dJ,Q+dQ)$. Then, by estimating the change in the minimum value, we can find the final state of the black hole: i) if the minimum value is zero, the final state is an extremal black hole of $(M+dM,J+dJ,Q+dQ)$ because there is still only one solution to the horizon; ii) if the minimum value is negative, the final state is a non-extremal black hole having two solutions corresponding to the inner and outer horizon; iii) if the minimum value is positive, there is no solution that has a horizon, and the final state is a naked singularity. These cases are shown in Fig.\,\ref{fig:fig1extremal}. Because the changes in the black hole are infinitesimally small, the first order of $\Delta_r(M+dM,J+dJ,Q+dQ)$ governs its physical behaviors. Therefore, 
\begin{align}
\Delta_r(M+dM,J+dJ,Q+dQ)&=\Delta_\text{min}+d\Delta_\text{min}\\
&=\frac{\partial \Delta_\text{min}}{\partial M}dM+\frac{\partial \Delta_\text{min}}{\partial J}dJ+\frac{\partial \Delta_\text{min}}{\partial Q}dQ+\frac{\partial \Delta_\text{min}}{\partial r_\text{min}}dr_\text{min}\nonumber\\
&=\frac{\partial \Delta_\text{min}}{\partial M}dM+\frac{\partial \Delta_\text{min}}{\partial J}dJ+\frac{\partial \Delta_\text{min}}{\partial Q}dQ,\nonumber
\end{align}
where we impose the extremal condition in Eq.\,(\ref{eq:extremalcond01}). In addition,
\begin{align}
\frac{\partial \Delta_\text{min}}{\partial M}=-\frac{2J^2}{M^3}-2\left(1+\frac{Q^2}{2M^2}\right)r_\text{min},\quad \frac{\partial \Delta_\text{min}}{\partial J}=\frac{2J}{M^2},\quad \frac{\partial \Delta_\text{min}}{\partial Q}=\frac{2Q r_\text{min}}{M}.
\end{align}
In combination with Eq.\,(\ref{eq:variationbh01}), the change in the minimum value is obtained as
\begin{align}
d\Delta_\text{min}=-8\pi^2 M r_\text{min} (\omega-m\Omega_\text{min}-q\Phi_\text{min})^2<0.
\end{align}
Therefore, the minimum value of the final black hole is always negative under the scattering of the scalar field, which corresponds to the case of the blue dashed line in Fig.\,\ref{fig:fig1extremal}\,(b). Consequently, the extremal black hole becomes a non-extremal black hole having two horizons. This implies that the extremal black hole cannot be over-spun or over-charged, and it can easily decay to a non-extremal black hole with a small perturbation of a scalar field. Hence, the WCCC is valid under the scattering for the extremal Kerr--Sen black hole.

\subsection{Near-Extremal Black Holes}

Herein, we generalize our discussion in Sec.\,\ref{sec41} to the case of near-extremal Kerr--Sen black holes. The near-extremal black hole is included in the case of non-extremal black holes, but it has a possibility of being over-spun or over-charged. To represent the near-extremal condition, we have to introduce an additional constant such that the constant can compete with the conserved quantities carried by the scalar field. If the carried conserved quantities are sufficiently large compared with the constant to enable the over-spinning or over-charging of the black hole, then the black hole can be a naked singularity. Such a situation has already been found in \cite{Siahaan:2015ljs} through testing with a particle, and the invalidity of the WCCC was resolved in \cite{Gwak:2016gwj} by considering the second-order perturbation due to the particle. Hence, in the present study, we investigate whether the WCCC can be invalid and whether we still need the second-order analysis as in \cite{Gwak:2016gwj} to resolve the invalidity.

Near-extremal black holes satisfy
\begin{align}
r_\text{h}>r_\text{min},\quad \Delta_\text{min}=\delta<0,\quad \frac{\partial \Delta_\text{min}}{\partial r_\text{min}}=0,\quad \frac{\partial^2 \Delta_\text{min}}{\partial r_\text{min}^2}>0,
\end{align}
where we introduce a negative constant $\delta$ to represent the near-extremality of the black hole. Furthermore, because near-extremal black holes are very close to being extremal black holes, we assume that $|\delta|\ll 1$. Note that the outer horizon is located at the front of the minimum in the near-extremal black hole. Then, the change in the minimum value is investigated under the scattering. Particularly, the sign of the changed minimum value is important to obtain the final state of the black hole, as we have done in Sec.\,\ref{sec41} and as shown in Fig.\,\ref{fig:fig1extremal}. The minimum value in the final state also depends on $\delta$.
\begin{align}
\Delta_\text{min}+d\Delta_\text{min}&=\delta+\frac{\partial \Delta_\text{min}}{\partial M}dM+\frac{\partial \Delta_\text{min}}{\partial J}dJ+\frac{\partial \Delta_\text{min}}{\partial Q}dQ+\frac{\partial \Delta_\text{min}}{\partial r_\text{min}}dr_\text{min}\\
&=\delta+\frac{\partial \Delta_\text{min}}{\partial M}dM+\frac{\partial \Delta_\text{min}}{\partial J}dJ+\frac{\partial \Delta_\text{min}}{\partial Q}dQ.\nonumber
\end{align}
Because the location of the minimum $r_\text{min}$ is very close to the outer horizon, we can rewrite $r_\text{min}$ in terms of $r_\text{h}$ by introducing $\epsilon$.
\begin{align}
r_\text{h}-r_\text{min}=\epsilon>0,\quad \epsilon\ll 1.
\end{align}
Hence, we can express the change in the minimum value based on the order of $\epsilon$.
\begin{align}
\frac{\partial \Delta_\text{min}}{\partial M}=-4r_\text{h}+4\epsilon +\frac{2\epsilon^2}{M},\quad \frac{\partial \Delta_\text{min}}{\partial J}=4r_\text{h}\Omega_\text{h},\quad \frac{\partial \Delta_\text{min}}{\partial Q}=4 r_\text{h}\Phi_\text{h}-4\epsilon \Phi_\text{h},\quad \delta=-\epsilon^2.
\end{align}
In combination with Eq.\,(\ref{eq:variationbh01}), the change in the minimum value is obtained as
\begin{align}
\Delta_\text{min}+d\Delta_\text{min}=&-8\pi^2 M r_\text{h} (\omega -m \Omega_\text{h}-q\Phi_\text{h})^2dt+8\pi^2 M r_\text{h} (\omega -m \Omega_\text{h}-q\Phi_\text{h})(\omega -q\Phi_\text{h})\epsilon dt+\mathcal{O}(\epsilon^2).
\end{align}
Changes in the black hole originate from conserved quantities carried by the fluxes of the scalar field during an infinitesimal time interval $dt$. Because the time interval is infinitesimal, its scale is the same as that of $\epsilon$, $dt\sim\epsilon$. Hence, only the first-order term remains.
\begin{align}\label{eq:}
\Delta_\text{min}+d\Delta_\text{min}=&-8\pi^2 M r_\text{h} (\omega -m \Omega_\text{h}-q\Phi_\text{h})^2dt<0.
\end{align}
The first-order term is always negative. Hence, the near-extremal black hole is stable under the scattering. Moreover, it cannot be over-spun or over-charged under the process. Therefore, the WCCC is valid for the initially near-extremal black hole. The validity of the WCCC of the near-extremal black hole in the first order is very interesting because it has not been reported in previous papers\cite{Jiang:2019vww,Duztas:2018adf}. When we consider the charged scalar field scattered by the Kerr--Sen black hole, the WCCC is valid in the first order of variations originating from the fluxes of the scalar field. Particularly, although the WCCC was found to be invalid in a neutral scalar field in \cite{Duztas:2018adf}, we resolve it by adding an electric charge to the scalar field. Furthermore, by considering the detailed fluxes of the charged scalar field, the WCCC is valid in the first-order variation, which is a leading order compared to the second order in \cite{Jiang:2019vww}. Note that the WCCC in Kerr--Sen black holes can be compared with that in Kerr--Newman black holes. The rotating parts of the two black holes are coincident to Kerr black holes, but their electric parts are different. The WCCC in Kerr black holes is known to be valid\cite{Gwak:2018akg}. The WCCC in Kerr--Newman or Reissner--Nordstr\"{o}m black holes is also valid\cite{Toth:2011ab}. Hence, although the metric components of the Kerr--Sen and Kerr--Newman black holes are different, the validity of WCCC is coincident in the two black holes.

\section{Laws of Thermodynamics under Charged Scalar Field}\label{sec5}

In Sec.\,\ref{sec4}, we have shown the validity of the WCCC with a charged scalar field. Furthermore, with the scattering of the scalar field, the extremality of extremal and near-extremal black holes decreases because the minimum value always becomes more negative. This implies that non-extremal black holes cannot evolve into extremal black holes under an external field. The extremal black hole also behaves like an unstable point in the states of the black hole. These phenomena may be quite similar to the third law of thermodynamics. Herein, we discuss and attempt to understand such behaviors and the WCCC in terms of the laws of thermodynamics.

\subsection{First and Second Laws} 

Most thermodynamic properties of a black hole are based on its outer horizon. Furthermore, the Hawking temperature and Bekenstein--Hawking entropy are defined at the location of the outer horizon. Because all conserved quantities of the black hole vary according to the fluxes of the charged scalar field, the location of the outer horizon also changes. Subsequent to the changes, the balance among thermodynamic properties can vary. However, because the variation originates from the external scalar field, it is unclear whether thermodynamic balance is maintained under this process.

The fluxes of the charged scalar field change the location of the black hole during a time interval. We have already identified that the final state of an initially extremal or near-extremal black hole is a non-extremal black hole. Hence, for any initial black hole, the final state should have an outer horizon. Thus, $\Delta_r$ always has a solution corresponding to the outer horizon.
\begin{align}\label{eq:outereqs01}
\Delta_r(M+dM,J+dJ,Q+dQ,r_\text{h}+dr_\text{h})=\frac{\partial \Delta_\text{h}}{\partial M}dM+\frac{\partial \Delta_\text{h}}{\partial J}dJ+\frac{\partial \Delta_\text{h}}{\partial Q}dQ+\frac{\partial \Delta_\text{h}}{\partial r_\text{h}}dr_\text{h}=0,
\end{align}
where
\begin{align}
\Delta_\text{h}&\equiv \left.\Delta\right|_{r=r_\text{h}}=0,\quad \frac{\partial \Delta_\text{h}}{\partial M}=-\frac{2a^2}{M}-2\left(1+\frac{Q^2}{2M^2}\right)r_\text{h},\nonumber\\
\frac{\partial \Delta_\text{h}}{\partial J}&=\frac{2a}{M}, \quad \frac{\partial \Delta_\text{h}}{\partial Q}=\frac{2Qr_\text{h}}{M},\quad \frac{\partial \Delta_\text{h}}{\partial r_\text{h}}=-2\left(M-\frac{Q^2}{2M}\right)+2r_\text{h}.\nonumber
\end{align}
The above expression implies that an outer horizon exists, even if conserved quantities change by as much as the quantities carried by the scalar field do. By solving Eq.\,(\ref{eq:outereqs01}) with Eq.\,(\ref{eq:variationbh01}), we can obtain a charge at the location of the outer horizon in terms of the variables of the scalar field.
\begin{align}\label{eq:changeinouterhorizon01}
dr_\text{h}=\frac{2dt \pi^2 r_\text{h} (am+qQr_\text{h}-(a^2+(b+M)r_\text{h})\omega)(-q\Phi_\text{h}+\omega-m\Omega_\text{h})}{-b+M-r_\text{h}},
\end{align}
which depends on the characteristics of the scalar field and can be larger or smaller along the modes of the scalar field. This result is quite different from what is known for Kerr--de Sitter black holes in Einstein's gravity\cite{Gwak:2018akg}. Because the distance of the outer horizon is closely related to the Bekenstein--Hawking entropy or the surface area of the black hole, we are concerned about the second law of thermodynamics. However, when Kerr--Sen black holes of $(M,J,Q,r_\text{h})$ change into black holes of $(M+dM,J+dJ,Q+dQ,r_\text{h}+dr_\text{h})$, the change in the entropy is obtained by combining Eqs.\,(\ref{eq:variationbh01}) and (\ref{eq:changeinouterhorizon01}) as
\begin{align}\label{eq:changeBHentropy01}
d S_\text{h}&=\frac{\partial S_\text{h}}{\partial M}dM+\frac{\partial S_\text{h}}{\partial J}dJ+\frac{\partial S_\text{h}}{\partial Q}dQ+\frac{\partial S_\text{h}}{\partial r_\text{h}}dr_\text{h}\\
&=\pi d(r_\text{h}^2+a^2+2br_\text{h})=\frac{16\pi^3 M^2 r_\text{h}^2(\omega -m\Omega_\text{h}-q\Phi_\text{h})^2}{d\Delta_\text{h}}dt,\nonumber
\end{align}
which is always positive. Hence, the entropy can be irreducible for any mode of the scalar field. This result is still consistent with the second law of thermodynamics. We can now rewrite the change in the mass of the black hole  of Eq.\,(\ref{eq:variationbh01}) in terms of other thermodynamic variables.
\begin{align}\label{eq:firstlaw1}
dM=T_\text{h}dS_\text{h}+\Omega_\text{h}dJ+\Phi_\text{h}dQ_\text{B}.
\end{align}
The above equation is exactly the first law of thermodynamics. Although the charged scalar field obeys its scalar field equations given by Eq.\,(\ref{eq:equationofmotion1}), rather than the equations of motion in the heterotic string theory, the conserved quantities carried by the scalar field vary those of the black hole according to the first law of thermodynamics in the first-order variation. From Eq.\,(\ref{eq:firstlaw1}), we also expect that the energy conservation is well defined in the scattering process. As in Kerr--Sen black holes, the first and second laws of thermodynamics are also well obtained in Kerr--Newman black holes by the scattering of the scalar field\cite{Gwak:2018akg,Gwak:2018tmy}.

\subsection{Third Law} 

\begin{figure}[h]
\centering
\subfigure[{For $\omega= 1$, $q= 1$, $m= 1$.}] {\includegraphics[scale=0.5,keepaspectratio]{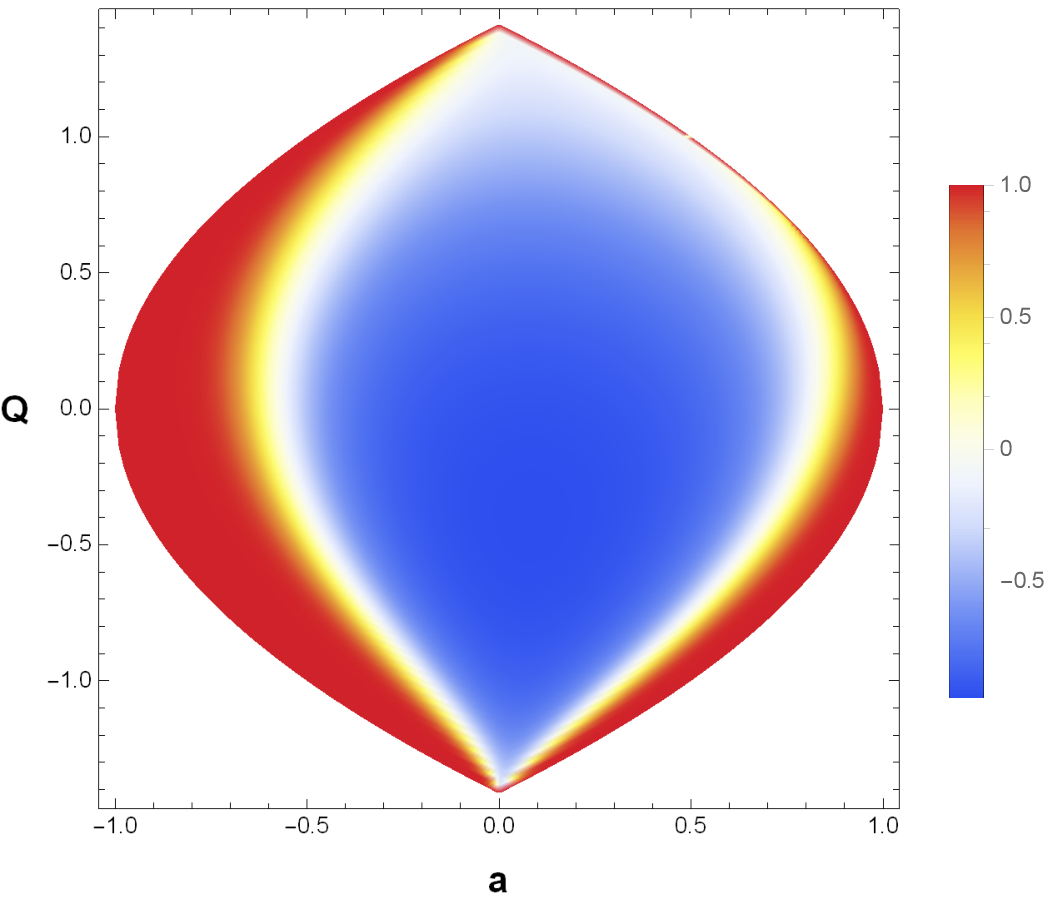}}\quad
\subfigure[{For $\omega= 1$, $q= 1$, $m= 3$.}] {\includegraphics[scale=0.5,keepaspectratio]{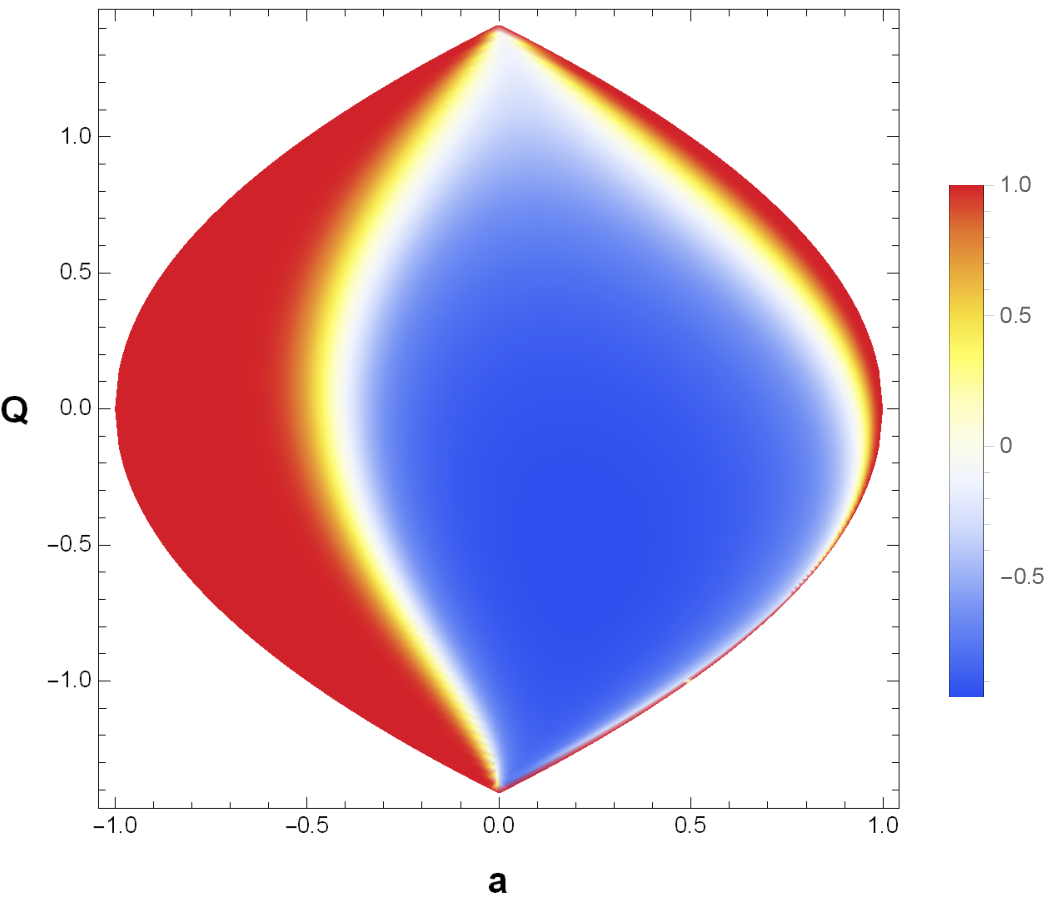}}\quad
\subfigure[{For $\omega= 1$, $q= 3$, $m= 1$.}] {\includegraphics[scale=0.5,keepaspectratio]{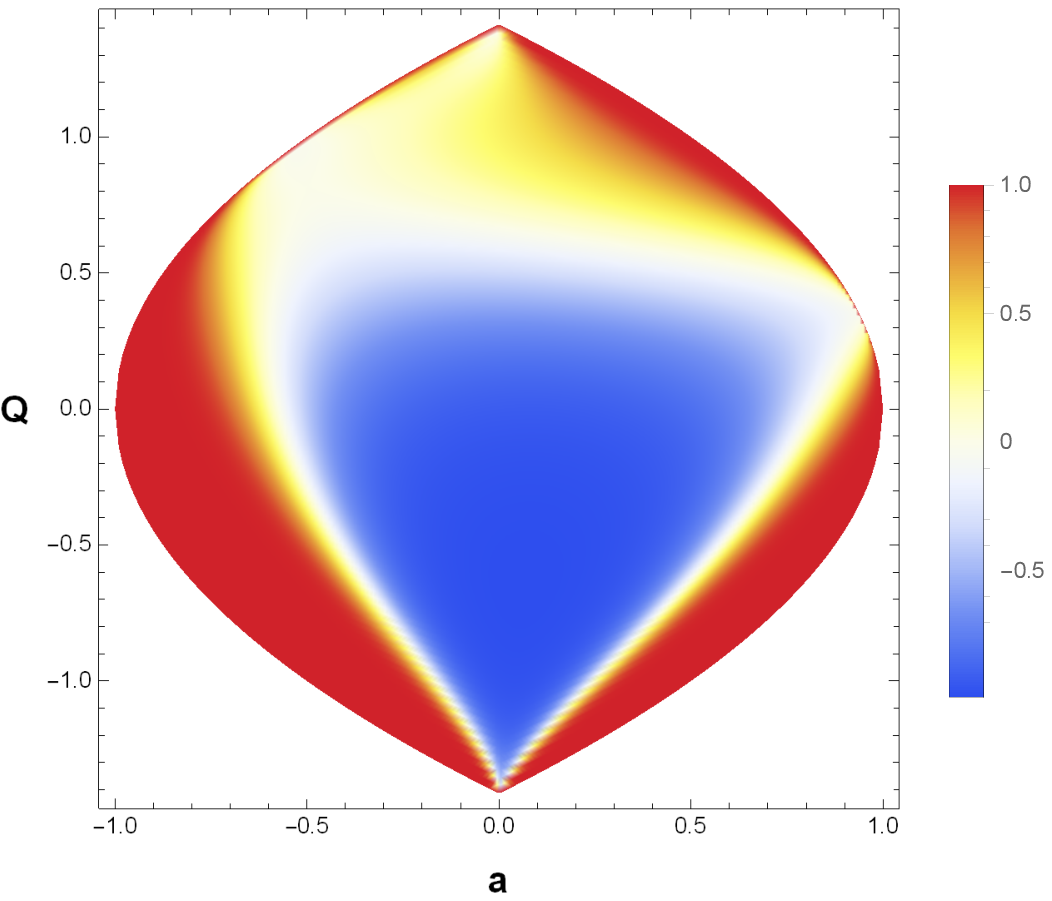}}
\caption{{\small Diagram for $\tanh(dT_\text{h}/dt)$ in Kerr-Sen black holes scattering a given charged scalar field.}}
\label{fig:fig2}
\end{figure}

The third law of thermodynamics states that zero temperature cannot be achieved by a physical process. As per the third law, because the extremal black hole is at zero temperature, a non-extremal black hole should not become an extremal black hole under the scattering of a charged scalar field. Under the scattering, the change in the Hawking temperature is obtained as
\begin{align}\label{eq:generaltemp01}
dT_\text{h}&=\frac{\partial T_\text{h}}{\partial M}dM+\frac{\partial T_\text{h}}{\partial J}dJ+\frac{\partial T_\text{h}}{\partial Q}dQ+\frac{\partial T_\text{h}}{\partial r_\text{h}}dr_\text{h}\\
&=\frac{\pi (\omega-m\Omega_\text{h}-q\Phi_\text{h})((a^2-b^2+M^2)\omega-am+(b-M)qQ)dt}{d\Delta_\text{h}M}-\pi (\omega-m\Omega_\text{h}-q\Phi_\text{h})^2dt,\nonumber
\end{align}
where the temperature can increase or decrease along the scalar field. However, if it is difficult for non-extremal black holes to become extremal black holes, the temperature of a near-extremal black hole can increase for any scalar field. Thus, we impose the near-extremal condition to Eq.\,(\ref{eq:generaltemp01}). The near-extremal condition for the outer horizon is
\begin{align}
\frac{\partial \Delta_\text{h}}{\partial r_\text{h}}=\zeta\ll1.
\end{align}
For near-extremal black holes, the change in the outer horizon is modified from Eq.\,(\ref{eq:changeinouterhorizon01}).
\begin{align}\label{eq:radiusnearextremal01}
dr_\text{h}=\frac{8\pi^2 M r_\text{h}^2}{\zeta}(\omega-m\Omega_\text{h}-q\Phi_\text{h})^2dt -2\pi^2 r_\text{h}^2 \omega (\omega-m\Omega_\text{h}-q\Phi_\text{h}),
\end{align}
where the leading term is more important and the outer horizon is large in the near-extremal black hole. By combining Eqs.\,(\ref{eq:generaltemp01}) and (\ref{eq:radiusnearextremal01}), the change in the temperature of the near-extremal black hole can be expressed as 
\begin{align}\label{eq:neartemperatureleading01}
dT_\text{h}=&2\pi r_\text{h} (\omega-m\Omega_\text{h}-q\Phi_\text{h})^2dt \zeta^{-1}+\pi (\omega-m\Omega_\text{h}-q\Phi_\text{h})\left(-\frac{r_\text{h}}{M}+m\Omega_\text{h}+2q \Phi_\text{h}\right)dt\\
&+\frac{\pi\omega(\omega-m\Omega_\text{h}-q\Phi_\text{h})}{4M}dt\zeta.\nonumber
\end{align}
Hence, when a black hole is close to the extremal state, the leading term in Eq.\,(\ref{eq:neartemperatureleading01}) is dominant; therefore, the temperature of the near-extremal black hole should increase. This result is numerically shown in Fig.\,\ref{fig:fig2}.
In the figure, changes to the normalized temperature are shown for various Kerr--Sen black holes with $M=1$. The boundaries of the graphs represent extremal black holes. Furthermore, the normalization preserves the signs of changes. Here, we find that the changes of near-extremal black holes are all positive. Therefore, non-extremal black holes cannot evolve into extremal black holes under such scattering. Consequently, we conclude that the third law of thermodynamics is also valid. This is important in the WCCC because the extremal black hole functions as a type of boundary in the evolution of the black hole. Therefore, the black hole cannot be over-spun or over-charged by the scattering of the charged scalar field.

\section{Summary}\label{sec6}

We investigated the validity of the WCCC for the Kerr--Sen black hole, which is a solution to the four-dimensional low-energy effective field theory for the heterotic string theory, based on the scattering of a charged scalar field. When the scalar field is scattered by the black hole, its conserved quantities, such as energy, angular momentum, and electric charge, flow into the black hole. We found the conserved quantities flowing into the black hole through the fluxes of the scalar field, and we estimated changes in the black hole during an infinitesimal time interval. The scattering changed all the variables of the black hole, such as mass, angular momentum, and electric charge, which determine the state of the black hole. Under the process, interestingly, the outer horizon stably exists for the charged scalar field with any mode, even if we assumed an initially extremal or near-extremal black hole. From the estimation, we found that the black hole cannot be over-spun or over-charged, and the singularity is stably hidden by horizons. Therefore, the WCCC is valid under the scattering.

We also studied the validity of the WCCC from a thermodynamic point of view. The first law of thermodynamics could be constructed for changes in the black holes due to the charged scalar field because energies such as rotational and electric energies are well balanced and conserved in the scattering process. Although the location of the outer horizon depends on the scalar field, the entropy of the scalar field is irreducible for any mode of the scalar field, which is one of reasons for the validity of the WCCC. This implies that the black hole cannot be shrunk. The other reason for the validity is the third law of thermodynamics. The temperature depends on the mode of the scalar field in a non-extremal black hole. However, the temperatures of near-extremal black holes increase for any mode of the scalar field. Consequently, the extremal black hole cannot reach zero temperature through such a physical process, and the black hole cannot be over-spun or over-charged. Thus, we confirmed that the WCCC is consistent with the laws of thermodynamics. Furthermore, the thermodynamic behaviors are closely related to the WCCC of the black hole.

\vspace{10pt} 

\noindent{\bf Acknowledgments}

\noindent This work was supported by the National Research Foundation of Korea (NRF) grant funded by the Korea government (MSIT) (NRF-2018R1C1B6004349) and the Dongguk University Research Fund of 2020.\\

\end{document}